# AN INNOVATIVE PLATFORM TO IMPROVE THE PERFORMANCE OF EXACT-STRING-MATCHING ALGORITHMS

Mosleh M. Abu-Alhaj[1], M. Halaiyqah[2], Muhannad A. Abu-Hashem[2], Adnan A. Hnaif[1], O. Abouabdalla[1,] and Ahmed M. Manasrah.

[1]: National Advanced IPv6 Center of Excellence, [2]: Computer Science
University Sains Malaysia, Penang Malaysia

## ABSTRACT

Exact-String-Matching is an essential issue in many computer science applications. Unfortunately, the performance of Exact-String-Matching algorithms, namely, executing time, does not address the needs of these applications. This paper proposes a general platform for improving the existing Exact-String-Matching algorithms executing time, called the PXSMAlg platform. The function of this platform is to parallelize the Exact-String-Matching algorithms using the MPI model over the Master/Slaves paradigms. The PXSMAlg platform parallelization process is done by dividing the Text into several parts and working on these parts simultaneously. This improves the executing time of the Exact-String-Matching algorithms. We have simulated the PXSMAlg platform in order to show its competence, through applying the Quick Search algorithm on the PXSMAlg platform. The simulation result showed significant improvement in the Quick Search executing time, and therefore extreme competence in the PXSMAlg platform.

## I. INTRODUCTION

Computer science applications play a significant role in many fields, such as DNA analysis, artificial intelligence, and information retrieval, among various others. String matching is an important issue in many of these applications. It is the process of finding the occurrence of a Pattern P into a Text T, wherein T is longer than P. This occurrence is either exactly matched or partially matched with the Pattern. Accordingly, string matching algorithms are divided into two main categories: Exact-String-Matching algorithms and approximate string matching algorithms. Exact-string-matching algorithms are concerned with the number of occurrences of the pattern into a given text, while approximate string matching algorithms are concerned with the similarity percentage between the pattern and the text or any part of the text [1] [2]. This paper concentrates on Exact-String-Matching algorithms, such as the Boyer-Moore, Horspool, and Quick Search algorithms [3].

Currently, the world is witnessing a revolution in hardware efficiency, where a normal laptop can have a multi-core processor. To take advantage of this revolution, most of the applications are used in parallel computing, wherein a problem is divided into smaller problems, which are then processed simultaneously. Moreover, many parallel paradigms and models have been developed and proposed. The Master/Slave paradigm is a widely used paradigm in parallel computing. It is a Multi-Processors paradigm containing several nodes, one node is the master and the other nodes are the slaves. The master node is responsible for maintaining global data structures and partitioning the overall computational problem into smaller sub–problems, which are handed to the slaves to process for computation. On the other hand, the Message Passing Interface (MPI) is one of the well-known parallel models used in parallel computing above the hardware and memory architectures. In this paper, we will use the MPI model along with the Master/Slave paradigm to develop a general parallel platform and improve the Exact-String-Matching algorithms' performance [4] [5] [6].

### I.1. Quick search algorithm

Sunday [7] proposed and designed a new algorithm for string matching, which is faster than the Boyer-moor algorithm and is considered one of the fastest algorithms in the string matching field. Its time and space complexity are $O(m + n)$ and $O(n)$, respectively. In terms of detecting matches between two strings, the quick search algorithm looks similar to the Boyer-moor algorithm. However, the difference between them is that the quick search algorithm only uses the bad-character shift table while the Boyer-Moore uses both bad-character shift and good suffix shift tables. Moreover, this algorithm starts searching from the left-most character to the right [7].

The rest of this paper is arranged as follows. Section 2 discusses some of the related works. Section 3 discusses the proposed platform, highlights the border problem, and shows the proposed platform performance. Finally, the conclusion is stated in Section 4.

## II. RELATED WORKS

There have been several research works on parallel Extract-String-Matching algorithms. For example, Raju and Babu [8] proposed a parallel technique for string matching algorithm. They considered the linear array with a reconfigurable pipelined bus system (LARPBS) and 2D LARPBS for string matching in their work, which has many existing applications such as cellular automata, computational biology, and string database. The proposed method introduced increases the speedup of the string matching process using LARPBS. They obtained time complexity O (1) for the string matching on 2D LARPBS where no preprocessing is done to the text and the pattern [8].

Park and George [9] presented a dataflow schemes string matching algorithms parallelization. In their work, they covered exact matching and k-mismatched problems, which they consider as sub-problems in the string matching field. The time complexity of the proposed parallel algorithm was $O((n/d)+\alpha)$, $0 \leq \alpha \leq m$, where n and m are the length of the text and pattern with (n >> m) and d is the number of streams used. The parallelism degree can be controlled by changing the value of the variable d, which is present in the input streams. Due to the one-pass dataflow algorithms, there was no preprocessing and memory space used for this schema [9].

## III. PARALLEL-EXACT-STRINGS-MATCHING-ALGORITHM

Exact-String-Matching is one of the main problems in many computer applications. One of the Exact-String-Matching problems is the slow matching process between the Pattern and the Text. Parallel computing is a key technique used to reduce the time of the Exact-String-Matching process. In this paper, we have exploited one of the Parallel computing models, namely, the MPI model, in order to provide a general platform to parallelize the Exact-String-Matching algorithms. The proposed platform, called Parallel-Exact-Strings-Matching algorithm (PXSMAlg), can be applied in all the Exact-String-Matching algorithms, such as Quick Search. The PXSMAlg platform has been developed to run the Master/Slave paradigm. [3] [5] [6].

### III.1. The PXSMAlg Platform Process

The parallelization process of the PXSMAlg platform is accomplished through a set of steps. First, the Master node reads the Pattern and the Text (Source-File). Second, the Master node calculates the Source-File size and divides it into multiple parts, according to the determined nodes number. Then, the Master node distributes each part to a specific node. After that, the searching function starts in each node to find the Pattern, with each node searching in its source file part. Before the final step is done, each node checks the border of its neighbor, except for the last node. The border issue will discussed later. Finally, the number of matches is collected from all nodes. Meanwhile, the Master node calculates all the collected results and then prints the total result. Figure 1 illustrates the PXSMAlg platform.

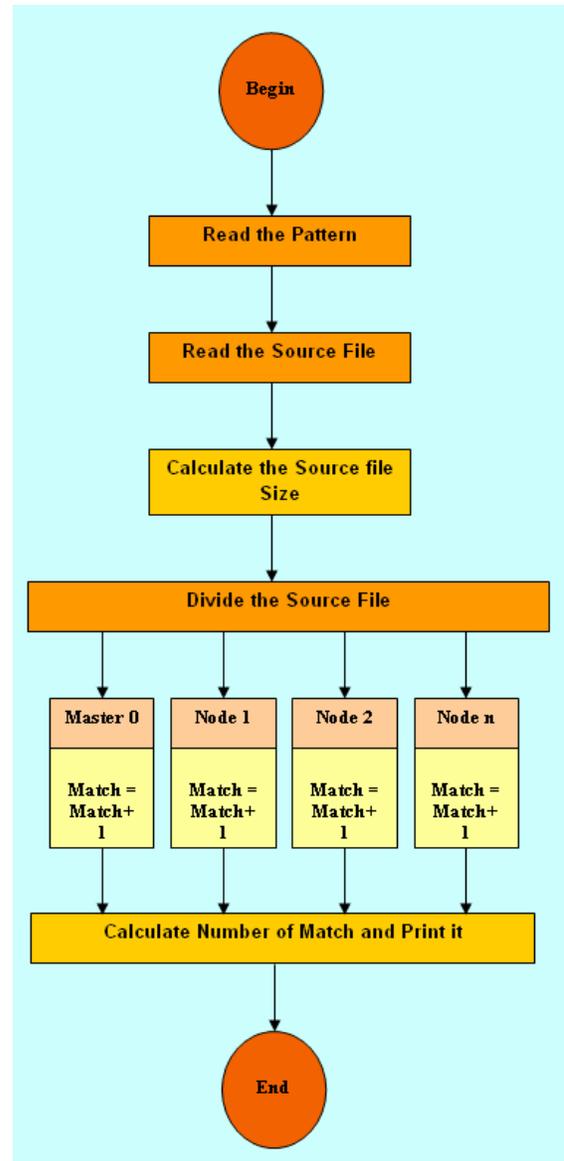

**Figure 1: The PXSMAlg platform process**

### III.2. Handling the Nodes Borders Issue

Border issue is key element that the PXSMAlg platform faces. This issue happened when the Pattern located between the Source-File parts borders led to a mismatched (i.e., not found) Pattern. To resolve this issue, the PXSMAlg platform allows node "n" to check the border between node "n" and node "n+1," node "n+1" to check the border between node "n+1" and node "n+2," and so on. For illustrative purposes, suppose the Source-File is "EXACT STRINGS MATCHING," the Pattern is "INGS," and the number of nodes is two. First, divide the Source-File into two parts according to the number of nodes, Part1 is "EXACT STRIN" and Part2 is "GS MATCHING." As we can notice, if node1 searched for the Pattern "INGS" in the border of the two parts, it will find it; otherwise, node1 will not be able to find the Pattern "INGS" in Part1 and node2 will not be able to find the Pattern "INGS" in Part2.

### III.3. PXSMAlg Platform Performance Analysis

We have built a simulation to demonstrate the feasibility of the PXSMAlg platform and its compatibility with the Exact-String-Matching algorithms. In addition, this simulation is done to compare the performance of the PXSMAlg platform with the conventional method, that is, the sequential method. The simulation built is based on three main factors: executing time, speedup, and efficiency. Our simulation runs under the Aurora server, which consists of 14 nodes, with each node having 2 CPUs, a speed of 1300MHz and a 1GB memory; all nodes run the Linux OS. The results showed high performance of the PXSMAlg platform over the sequential methods.

We have carried out 14 different experiments to search for the letter "a" in a 37 MB file size. We have applied the experiments using the Quick Search algorithm, which is one of the best algorithms in the Exact-Strings-Matching algorithms. The result showed significant improvement in the executing time and speedup, wherein applying the Exact-String-Matching algorithms on the PXSMAlg platform decreased the executing time, especially when compared with the sequential executing time. Figure 2 depicts the improvement in the Quick Search algorithm process time in the sequential mode, one node, parallel mode, and two or more nodes. In addition, the speedup is increased by applying the Exact-String-Matching algorithms on the PXSMAlg platform. Figure 3 shows the improvement in speedup in the Quick Search algorithm. In contrast to the executing time and the speedup, the processors' efficiency decreases by applying Exact-String-Matching algorithms on the PXSMAlg platform. Figure 4 shows the decreasing efficiency in the Quick Search algorithm when the number of the processors increases.

### IV. CONCLUSION

In this paper, we have proposed a general platform, called the PXSMAlg platform, in order to improve the Exact-Strings-Matching algorithms performance. The PXSMAlg platform relies on using the MPI model over the Master/Slave Paradigm to improve the Exact-Strings-Matching algorithms' competence in terms of speeding up the executing time. We have applied one of the best Exact-String-Matching algorithms, the Quick Search algorithm, on the PXSMAlg platform. The result showed high efficiency in the PXSMAlg platform. In comparison with the sequential mode, the Quick Search executing time and speedup were highly improved. On the other hand, the efficiency of the processors decreased.

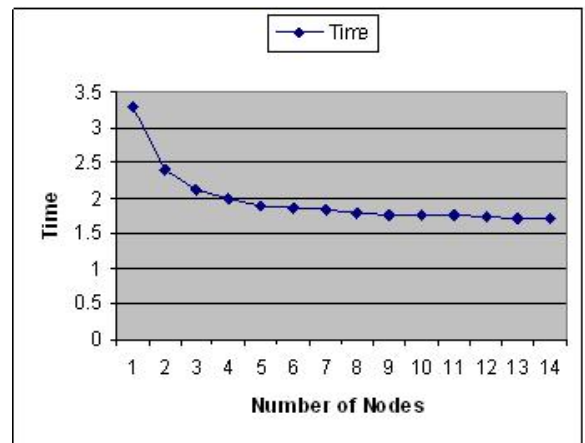

**Figure 2: Executing Time**

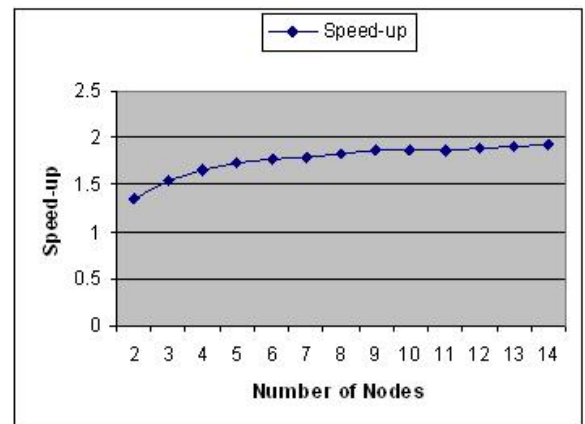

**Figure 3: Speedup**

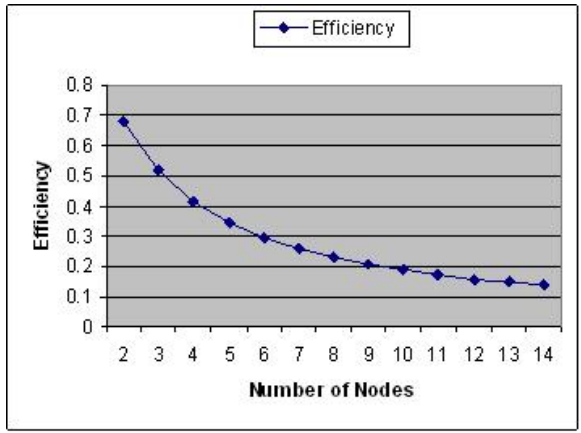

**Figure 4: Efficiency**

In Proceedings of the 32rd Hawaii International Conference on System Sciences, 1999.